# A Validated Method for Predicting Small Molecule Ionization Sites using Gibb's Free Energies

Jessica L. Bade[1], Sean M. Colby[1], Ryan S. Renslow[1,*], Thomas O. Metz[1,*],

[1]Earth and Biological Sciences Directorate, Pacific Northwest National Laboratory, Richland, Washington 99354, United States

*ryan.renslow@pnnl.gov; thomas.metz@pnnl.gov

**ABSTRACT:** Thorough and accurate molecular identification of unknown metabolites can unlock new areas of the molecular universe and allow greater insight into complex biological and environmental systems than currently possible. Analytical approaches for measuring the metabolome, such as nuclear magnetic resonance spectroscopy, and hyphenated separation techniques coupled with mass spectrometry, such as liquid chromatography-ion mobility spectrometry-mass spectrometry (LC-IMS-MS), have risen to this challenge by yielding rich experimental data that can be queried by cross-reference with similar information for known standards in appropriate reference libraries. Confident identification of molecules in metabolomics studies, though, is often limited by the diversity of available data across chemical space, the unavailability of authentic reference standards, and the corresponding lack of comprehensiveness of standard reference libraries. The In Silico Chemical Library Engine (ISiCLE) addresses the aforementioned hindrances by providing a first-principles, cheminformatics pipeline that yields collisional cross section (CCS) values for any given molecule and without the need for training data. In this program, chemical identifiers undergo molecular dynamics simulations, quantum chemical transformations, and ion mobility calculations for the generation of predicted CCS values. Here, we present a new module for ISiCLE that addresses the sensitivity of CCS predictions to ionization site location. An update to adduct creation methods is proposed concerning a transition from pKa and pKb led predictions to a Gibb's free energy (GFE) based determinacy of true ionization site location. A validation set of experimentally confirmed molecular protonation sites was assembled from literature and cross-referenced with the respective pKb predicted locations and GFE values for all potential ionization site placements. Upon evaluation of the two methods, the lowest GFE value, as based upon density functional theory calculations, was found to predict the true ionization site location with 100% accuracy while pKb had less accuracy. Furthermore, ionization site location was found to influence the corresponding calculated CCS values and is relevant to reducing the current error (3.2%) in ISiCLE's calculated versus experimental CCS. The impact of potential mis-selection of ionization site location on final molecular property prediction values signifies the importance of acknowledging distinct structural populations when interpreting experimental data, such as from LC-IMS-MS measurements, and attributing molecular characteristics.

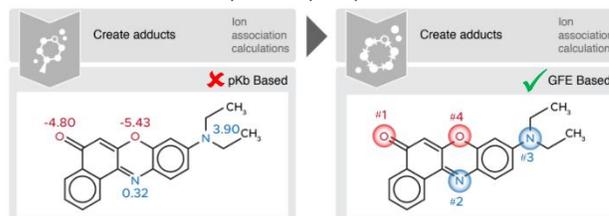

## Introduction

Small molecule chemical space is vast and estimated at more than $10^{60}$ predicted compounds[1-2]. Knowledge of the presence and potential bioactivities of those molecules in biological and environmental systems can be determined using metabolomics and is essential to improved understanding of human health, as well as the fields of exposomics and chemical forensics science. Particularly for the field of metabolomics, there is immense need for a capability that can provide comprehensive and unambiguous identification of chemical components in biological and environmental samples, as the majority of molecules are not available as authentic standards and thus not represented in identification libraries.

The metabolomics community has made extensive progress in decoding the "molecular universe" through the application of nuclear magnetic resonance (NMR)-structural elucidation and separations (e.g. ion mobility spectrometry; IMS) coupled with mass spectrometry (MS) to untargeted metabolomic profiling[3-4]. These and emerging analytical approaches can help remove molecular ambiguity in complex samples through confident identification of molecules through cross-reference with authentic reference standards. Experimental approaches, though, encounter the boundaries of science at a crossroad of high-throughput analysis and time-intensive structural elucidations. This forces a trade-off between sample complexity and further expansion into uncharted chemical space.

The most common techniques for high-throughput molecule identification involve cross-reference of results from GC-MS, LC-MS, and NMR studies with data from analyses of authentic reference standards[4]. The diversity and comprehensiveness of reliable reference standard data currently stands as a primary bottleneck to metabolomics with <1% of chemical space having a known standard[5]. Furthermore, the usage of standards confines metabolomics primarily to known and expected substances[6]. Through the use of standards-free approaches in coordination with bourgeoning high-dimensional analytical approaches, the farthest reaches of unknown, unexpected chemical space can be made accessible.

A supplemental approach is applying quantum chemical calculations to molecular property determination and cross-referencing predictions with appropriate experimental data[5]. We previously developed the In Silico Chemical Library Engine (ISiCLE) for generation of predicted molecular properties including collisional cross section (CCS) values, based on first principles. CCS is an experimentally identifiable molecular property that represents the apparent surface area of a molecular ion in $Å^2$.[7] This property has been calculated by ISiCLE to a 3.2% error for a set of nearly 2000 molecules whose experimental CCS were determined using multi-field IMS measurements[5]. CCS is derived from the drift time of ions separated by size, shape and charge as they pass through an electric field in the presence of an inert gas (e.g. $N_2$ and He) [7].

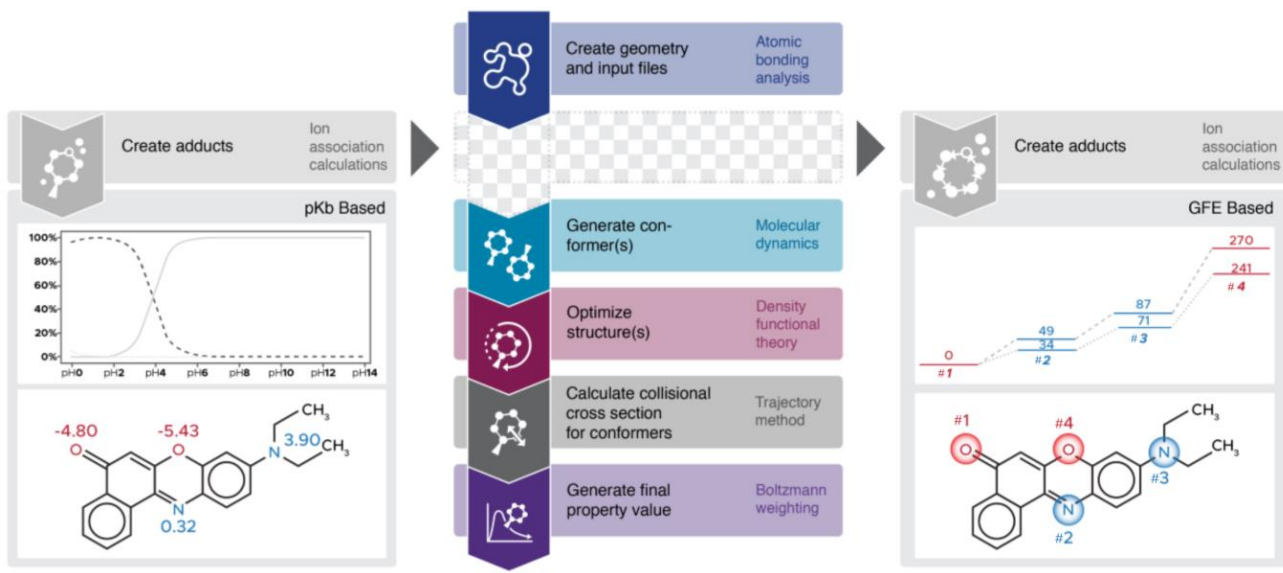

*Figure 1. ISiCLE workflow overview depicting changes in the adduct formation (ionization site prediction) step.* pKb predictions, left, are made by ChemAxon's Marvin software for oxazine dye Nile red. The most positive pKb value is indicative of the first location that would receive a proton. GFE calculated values (kJ/mol) are shown, right, for Nile red relative to the most negative energy value. Dotted line represents ISiCLE calculated values and dashed line represents published values. Atom site #1 is found the be the lowest relative energy location and confirmed true proton location in the gas phase via IRMPD spectroscopy[16].

Here, we have updated ISiCLE to include molecular adduct ionization site predictions based upon ion placement at every potential atom location and the removal of potential influence of a liquid phase. More specifically, the relevance of ionization site location with protomer populations is tested in this paper, and predictions are compared to experimental results. This application extends to the other supported adduct types in ISiCLE (e.g. deprotonation, sodiation). Another protonation prediction tool, part of the CREST software by the Grimme lab, is additionally surveyed[8]. An assembled set of molecules each with experimentally and computationally confirmed protonation sites are cross-referenced to assess the accuracy and sensitivity of the proposed methodology for adduct generation.

**pKa Theory.** Predicted pKa values are used in ion mobility prediction software of ISiCLE[5], MetCCS[9], and by Paglia et al.[10] While calculated pKa can offer a strong first approach to quickly resolving gas phase behaviours, e.g. ionization site placements, oversights in its application can lead to exclusion of relevant structural populations. There are two aspects of pKa applicability to consider: first, the relatability of predictions from different phasic environments, and second, the accuracy of third-party software predictions. Errors in computational chemistry amplify quickly and sometimes to a disastrous extent when considering the sensitivities of the tools used and the error bounds for scientific relevance. As such, a greater focus has been placed on the accuracy of pKa prediction tools to assess their relatability in a gas phase prediction pipeline.

The accuracy in pKa predictions of computational chemistry software can be significantly influenced by the chemical space representation of the training set. Balogh et al[11] tested the pKa prediction software ACD, Epik, Chemaxon Marvin (cxcalc), PharmaAlgorithm, and Pallas on an in-house data set of 95 measured compounds and additional challenging public compounds. These programs were found to predict with error exceeding 1 pKa (MAE) unit between 32% to 55% of the time and with poor coverage[11], meaning for some in-house molecules no pKa value was generated at all by the programs. While the field has advanced since 2012, large errors due to a small sampling highlights the importance of questioning the accuracy of third-party tools as well as the breadth of application.

Currently, the first pKa or pKb predicted site by Marvin's cxcalc[12] tool for negative or positive ionization, respectively, is treated as the only adduct site in the ISiCLE pipeline. The chance for potential mis-selection, non-selection, and overall molecular ambiguity due to the pKa and pKb application here is a reason to explore all atom sites and a source for improvement in small molecule identification as well as the ISiCLE framework.

## Computational Methods

The need to redress ionization ambiguity in ISiCLE became apparent in the process of updating software methods. A more expansive approach to ionization site placement testing is proposed here to prevent unjustified discrimination of potentially competitive ionization site locations. The modified adduct procedures and rationale are further explained here, while the reasoning for all other steps are discussed in a previous publication[5].

ISiCLE leverages a Snakemake[13] pipeline for its ability to seamlessly connect successive computational chemistry calculations. Snakemake is a Python-based workflow management system designed primarily to streamline scalability management and manage variable file naming schemes within informatics pipelines. Snakefiles, of the Snakemake language, are comprised of chained rules linked through filename variables within the input and output of every rule, where each rule can at a minimum run Python scripts or shell commands. A primary benefit of using a workflow management system is the standardization and hands-off involvement when running software, especially in high-throughput applications, as a single command is required to run all processes start to finish. This type of pipeline also enables a modular architecture, permitting alternative and parallel workflows within one program, requiring a sole download at GitHub (github.com/pnnl/isicle).

**Validation Set.** For this study, 11 molecules from literature with comparable experimental conditions were selected. Each molecule has experimentally confirmed adduct populations with established computational energies and order of ionization. These were all validated experimentally via methods including collision induced dissociation (CID) fragment pattern matching[14-16], infrared (multiple) photon dissociation (IR(M)PD)[17-19], or high field asymmetric waveform ion mobility spectrometry (FAIMS) analysis[20]. Discrepancies between validated molecules include the varying levels of density functional theory (DFT) and methods of experimental confirmation.

**Molecule Prep.** The initial steps of the ISiCLE pipeline involve parsing a file of InChI or SMILES strings to individual files queued in parallel through the program. For this paper, chemical identifiers were sourced from ChemSpider[21]. The workflow proceeds to initial chemical identifier preparation with molecule desalting, neutralization, and generation of a major tautomer.

**Adduct Formation.** The original adduct selection mechanism pulls input from the output of the previous Snakemake rule: geometry coordinates for the major tautomer. Marvin cxcalc pKa and pKb predictions are then generated for the tautomer at pH values in the range of -40 to 40. The atom sites associated with the lowest pKa and highest pKb value are selected as the only atom sites for ion placement.

The new adduct creation procedure of this study dubbed the Gibb's free energy, or GFE, method pulls in the same major tautomer geometry output. This adduct generation rule calls to a Python file that iterates over all atoms of the specified molecule and places a [-H], [+H], and [+Na] adduct at each atom where appropriate (i.e. [-H] at all hydrogen sites and [+H], [+Na] at all heavy atom, i.e., non-hydrogen, sites), and subsequently calculates the relative GFE of the adducts for each ionization site location (details below). For the purpose of direct comparisons with the validation set of molecules, only [M+H]$^+$ adducts are assessed in this paper.

Both methods of adduct generation are applied to the experimentally validated molecules, with the Universal Force Field [22](UFF) at 10 steps specified in the configuration file.

**Adduct Filtering.** Computational time can easily bloom to be intractable when using the Gibb's free energy method and when considering a molecule of X atoms × Y adducts × Z conformers results in X×Y×Z number of DFT calculations. To that end, two methods of adduct candidate filtering were considered. The first method assessed is a wider application of Chemaxon's cxcalc pKa tool. Instead of only considering the first identified atom by that tool, as is default, the possibility of testing all atoms identified out to a max of X specified atoms was evaluated. Alternatively, the xtb CREST program is also tested in this paper[23]. CREST offers a command-line protonation functionality that determines the most likely sites of protonation based off localized molecular orbitals calculations.

**Conformer Selection.** The ISiCLE workflow is shown in **Figure 1**. Conformer generation occurs pre-DFT and post any potential screening. ISiCLE queues up AMBER Molecular Dynamics after adduct selection for 10 default (set via configuration file) simulated annealing steps with 10 conformers selected from each step[24]. The 100 resultant conformers are down selected to the one most similar to the set and two least similar conformers of each annealing step, yielding 30 total conformers. Similarity is defined by the root-mean-squared deviation calculated from atomic-positions. The most similar conformer has the lowest pair-wise deviation sum in comparison to all other conformers in the annealing step and the two most dissimilar have the highest deviation sums.

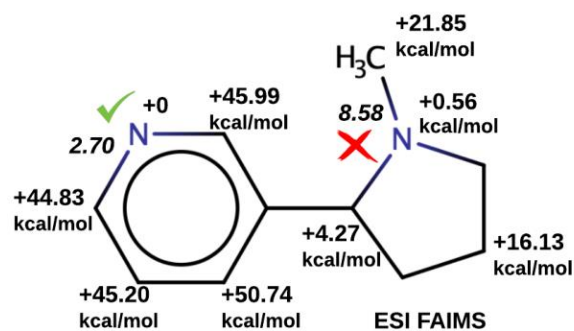

*Figure 2. Nicotine protonation locations with associated GFE values (kcal/mol) and pKb predictions (italics). The green check is indicative of the experimentally confirmed gas phase major protomer[18]. The red 'x' is indicative of the incorrectly predicted major protomer according to pKb calculations. All energy values are relative to the lowest energy value, denoted +0.*

**Density Functional Theory and CCS generation.** Molecular energy, geometry optimizations, and frequency calculations via NWChem's DFT methods are carried out post-conformer down-selection. NWChem is an open-source, computational chemistry software scalable to meet high-performance computing needs[25]. All molecules were optimized using the B3LYP exchange correlation functional. Basis sets were sourced from the Environmental Molecular Sciences Laboratory Basis Set Exchange[26]. Multiple Pople basis sets were considered to ascertain the minimum theory required for a confident identification of ionization order. ISiCLE's default DFT basis set of 6-31G* (a double-ζ valence potential basis set having a single polarization function) is applied in the remaining calculations. GFE calculations use NWChem output generated at 298K. Sample NWChem and structural inputs and calculations can be found in **SI1**, as well as basis set references.

**Post-DFT Processing.** The resultant geometry and energy optimized geometry files are propagated to our modified MOBCAL CCS trajectory method (MOBCAL-SHM), which is >2 orders of magnitude more computationally efficient than traditional MOBCAL, as described in earlier work[5]. CCS values for all conformers are ultimately Boltzmann weighted according to GFE values down to one representative CCS value.

## Results and Discussion

**Validation Set Comparison.** Despite the potential for variance of protomer populations between solvated and un-solvated environments, existing research on the evaluation of this discrepancy has thus far primarily occurred on a case by case basis. A literature search revealed 11 molecules with experimentally confirmed protonation locations in the gas phase, which provided the basis for more direct assessments of pKb predictions and GFE trends.

Two molecules, para-dimethylaminobenzoic acid[14] and uracil[19, 27], were evaluated with proton placement at every single non-hydrogen atom location and at varying levels of B3LYP DFT (**SI2, 3**). For these molecules, the validated ion location was found to be the same as the lowest GFE value ionization site placement across all DFT levels of theory. While the calculated energies may not have matched those calculated in the literature, the ordering of energies across atom locations trended the same for all levels of theory as confirmed in the source literature.

The remaining molecules of the validation set were also run through ISiCLE with both the dissociation constant- and GFE-based methods. As shown in **Fig. 1**, the ionization site of Nile Red is incorrectly predicted by pKb, whereas the third most favorable ionization site placement predicted to form by pKb is predicted by the GFE method as the most favorable site and experimentally confirmed as the only adduct formed. Nicotine, **Fig. 2**, is another example incorrectly predicted via the pKb method and with a smaller margin of error as the two most competitive ionization site locations are within 1 kcal/mol of each other. The ionization site locations for five out of the 11 molecules were found to be incorrectly predicted by pKb calculations, whereas 11 out of 11 were correctly predicted through GFE study (**Fig. 3**).

**Adduct Filtering.** Exhaustive ionization site location testing can compound to rather large computation times as seen in **SI4**. While the need for filtering is understandable as molecules increase in size, there is also a need to control for applicability of a filter to diverse sets of molecules. For the 11 molecules in the validation set, pKb prediction accuracy was mixed and for benzoic acid no location was identified at all (**Fig. 3, SI5**). Using pKa and pKb predictions for filtering when there may be no sites identified would be problematic to a high-throughput application.

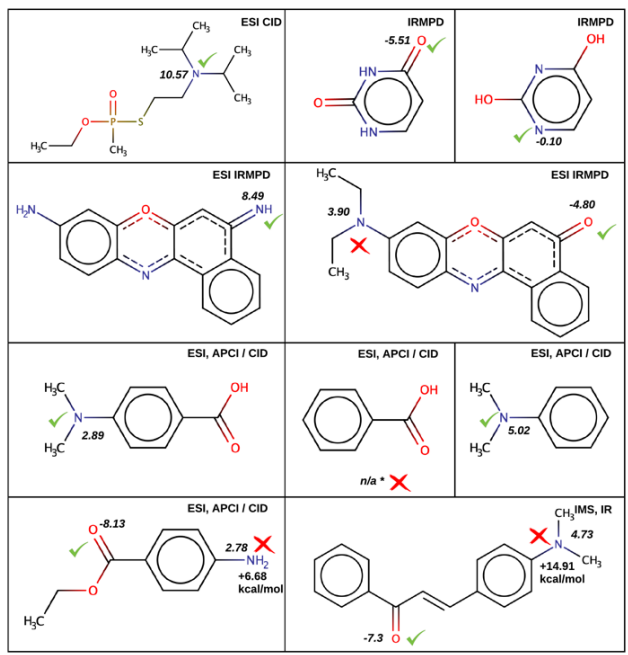

*Figure 3. Depiction of all molecules considered for validation. The method of validation is noted in the upper right corner of each box. A check indicates the correct protonation site. An 'x' represents an incorrectly predicted ionization location, as predicted by ChemAxon Marvin's cxcalc pKb predictor. If no 'x' is present, pKb accurately predicted the true ion location. All sites are correctly predicted via the GFE method. *No pKb prediction for this molecule.*

An alternative method of generating potential protomers is an application of the Grimme group's xtb CREST tool (v 2.7.1)[8, 23]. The protonation option of the command line program was also tested on the validation set of molecules. The tool quickly identifies a list of relevant protonated structures sorted according to energy values (GFN2-xTB theory). The tool was found to have 100% coverage of the correct ion locations present in the generated list for every molecule in the validation set. However, the experimentally confirmed adduct was not always the first (i.e., lowest energy) predicted molecule. This tool is an attractive candidate for filtering due to its complete coverage and speed.

**Effect on CCS.** Through analysis of optimized structures, their respective GFE, and the resultant CCS values generated, protomers were found to cluster to distinct energy levels and CCS ranges. One molecule of interest, 4-(dimethylamino)benzoic acid (seen in **SI6**) has a confirmed ionization site location with an associated predicted CCS far removed from the majority of other adducts' CCS values. While this molecule was accurately predicted by pKb, the next most likely site to form is 7-8 $Å^2$ away in CCS value, demonstrating the large effect that adduct ionization site location can have on CCS value predictions. An error in adduct mis-selection could have yielded an incorrect CCS and therefore an incorrect molecule identification in this case. Additionally, the work by Nielson et al. found conformational populations to exhibit similar behaviour to protomer populations[28]. They assert that an adequate sampling of conformational space is also necessary to identify the most prevalent conformer in the gas phase. The recommendation for future molecule gas phase predictions is now to thoroughly sample both relevant ionization site locations and conformational space.

The use of dissociation constant-based predictions for ionization site placement was compared to GFE-based ion placements for 25 molecules considered in the original ISiCLE paper that were found to have a larger error in CCS ([M+H], [M+Na], [M-H]) value predictions than expected. pKb predictions of ionization site placement were found to match GFE method prediction for <60% of these 25 problematic molecules (**SI7**), and in a few cases, failed to predict it at all (e.g., no pKb prediction returned). This suggested, as suspected, that the GFE and dissociation constant-based methods could yield differing most likely sites for ionization placement.

## Conclusions

The updated adduct prediction methods of ISiCLE for greater inclusion of potential ionization site locations met the purpose of this work for more accurate ionization location calculations. Through simple consideration of more ion locations with GFE weighting, it was found that a less discriminatory approach early-on can have significant impact further down the ISiCLE pipeline. Marvin cxcalc pKa/pKb based prediction methods were found to be <100% accurate in predicting the true ionization site location in both the first site and in all sites predicted. In comparison with the validated set of molecules, GFE identified ionization site locations were found to be 100% accurate in predicting the experimentally observed gas phase locations. While the validated set of molecules is small due to the limited literature available and applicability to ISiCLE, this sampling was enough to show GFE is a significant improvement for ion location predictions over computational pKa/pKb tools.

Furthermore, CCS is known to be highly susceptible to isomer changes and is found to be additionally influenced by charge location.[29] A more precise CCS prediction is essential to standards-free reference libraries generated through computational predictions in small molecule identification, as the main barrier to their usage is the known error. While there may be improved predictions for ISiCLE with a lower error, the goal of this paper is to increase awareness of the impact charge location could have on final property prediction in metabolomics.

Supporting Information. The Supporting Information is available free of charge on the ACS Publications website.

SupportingInformation.pdf: Includes further detailed methods and additional figures.

Corresponding Authors.

Ryan S. Renslow - ryan.renslow@pnnl.gov

Thomas O. Metz – thomas.metz@pnnl.gov

Author Contributions. The manuscript was written primarily by J.B. and through contributions of all other authors. All authors have given approval to the final version of the manuscript.

Acknowledgements. This work utilized resources and capabilities developed by the Pacific Northwest Advanced Compound Identification Core under National Institutes of Health, National Institute of Environmental Health Sciences grant U2CES030170.

## References

1. Bohacek, R. S.; McMartin, C.; Guida, W. C., The art and practice of structure-based drug design: A molecular modeling perspective. *Medicinal Research Reviews* **1996,** *16* (1), 3-50.

2. Reymond, J.-L.; Awale, M., Exploring Chemical Space for Drug Discovery Using the Chemical Universe Database. *ACS Chemical Neuroscience* **2012,** *3* (9), 649-657.

3. May, J. C.; Goodwin, C. R.; McLean, J. A., Gas-Phase Ion Mobility-Mass Spectrometry (IM-MS) and Tandem IM-MS/MS Strategies for Metabolism Studies and Metabolomics. *Encyclopedia of Drug Metabolism and Interactions* **2012**, 1-29.

4. Schrimpe-Rutledge, A. C.; Codreanu, S. G.; Sherrod, S. D.; McLean, J. A., Untargeted Metabolomics Strategies-Challenges and Emerging Directions. *Journal of the American Society for Mass Spectrometry* **2016,** *27* (12), 1897-1905.

5. Colby, S. M.; Thomas, D. G.; Nuñez, J. R.; Baxter, D. J.; Glaesemann, K. R.; Brown, J. M.; Pirrung, M. A.; Govind, N.; Teeguarden, J. G.; Metz, T. O.; Renslow, R. S., ISiCLE: A Quantum Chemistry Pipeline for Establishing in Silico Collision Cross Section Libraries. *Analytical Chemistry* **2019,** *91* (7), 4346-4356.

6. De Vijlder, T.; Valkenborg, D.; Lemière, F.; Romijn, E. P.; Laukens, K.; Cuyckens, F., A tutorial in small molecule identification via electrospray ionization-mass spectrometry: The practical art of structural elucidation. *Mass Spectrometry Reviews* **2018,** *37* (5), 607-629.

7. Lanucara, F.; Holman, S. W.; Gray, C. J.; Eyers, C. E., The power of ion mobility-mass spectrometry for structural characterization and the study of conformational dynamics. *Nature Chemistry* **2014,** *6* (4), 281-294.

8. Pracht, P.; Bohle, F.; Grimme, S., Automated exploration of the low-energy chemical space with fast quantum chemical methods. *Physical Chemistry Chemical Physics* **2020,** *22* (14), 7169-7192.

9. Zhou, Z.; Xiong, X.; Zhu, Z.-J., MetCCS predictor: a web server for predicting collision cross-section values of metabolites in ion mobility-mass spectrometry based metabolomics. *Bioinformatics* **2017,** *33* (14), 2235-2237.

10. Paglia, G.; Williams, J. P.; Menikarachchi, L.; Thompson, J. W.; Tyldesley-Worster, R.; Halldórsson, S.; Rolfsson, O.; Moseley, A.; Grant, D.; Langridge, J.; Palsson, B. O.; Astarita, G., Ion Mobility Derived Collision Cross Sections to Support Metabolomics Applications. *Analytical Chemistry* **2014,** *86* (8), 3985-3993.

11. Balogh, G. T.; Tarcsay, Á.; Keserű, G. M., Comparative evaluation of pKa prediction tools on a drug discovery dataset. *Journal of Pharmaceutical and Biomedical Analysis* **2012,** *67-68*, 63-70.

12. *cxcalc*, 16.11; ChemAxon: 2017.

13. Köster, J.; Rahmann, S., Snakemake—a scalable bioinformatics workflow engine. *Bioinformatics* **2012,** *28* (19), 2520-2522.

14. Chai, Y.; Weng, G.; Shen, S.; Sun, C.; Pan, Y., The Protonation Site of para-Dimethylaminobenzoic Acid Using Atmospheric Pressure Ionization Methods. *Journal of The American Society for Mass Spectrometry* **2015,** *26* (4), 668-676.

15. Housman, K. J.; Swift, A. T.; Oyler, J. M., Fragmentation Pathways and Structural Characterization of 14 Nerve Agent Compounds by Electrospray Ionization Tandem Mass Spectrometry. *Journal of Analytical Toxicology* **2014,** *39* (2), 96-105.

16. Warnke, S.; Seo, J.; Boschmans, J.; Sobott, F.; Scrivens, J. H.; Bleiholder, C.; Bowers, M. T.; Gewinner, S.; Schöllkopf, W.; Pagel, K.; von Helden, G., Protomers of Benzocaine: Solvent and Permittivity Dependence. *Journal of the American Chemical Society* **2015,** *137* (12), 4236-4242.

17. Chai, Y.; Hu, N.; Pan, Y., Kinetic and Thermodynamic Control of Protonation in Atmospheric Pressure Chemical Ionization. *Journal of The American Society for Mass Spectrometry* **2013,** *24* (7), 1097-1101.

18. Nieckarz, R. J.; Oomens, J.; Berden, G.; Sagulenko, P.; Zenobi, R., Infrared multiple photon dissociation (IRMPD) spectroscopy of oxazine dyes. *Physical Chemistry Chemical Physics* **2013,** *15* (14), 5049-5056.

19. Salpin, J.-Y.; Guillaumont, S.; Tortajada, J.; MacAleese, L.; Lemaire, J.; Maitre, P., Infrared Spectra of Protonated Uracil, Thymine and Cytosine. *ChemPhysChem* **2007,** *8* (15), 2235-2244.

20. Marlton, S. J. P.; McKinnon, B. I.; Ucur, B.; Maccarone, A. T.; Donald, W. A.; Blanksby, S. J.; Trevitt, A. J., Selecting and identifying gas-phase protonation isomers of nicotineH+ using combined laser, ion mobility and mass spectrometry techniques. *Faraday Discussions* **2019,** *217* (0), 453-475.

21. Pence, H. E.; Williams, A., ChemSpider: An Online Chemical Information Resource. *Journal of Chemical Education* **2010,** *87* (11), 1123-1124.

22. Rappe, A. K.; Casewit, C. J.; Colwell, K. S.; Goddard, W. A.; Skiff, W. M., UFF, a full periodic table force field for molecular mechanics and molecular dynamics simulations. *Journal of the American Chemical Society* **1992,** *114* (25), 10024-10035.

23. Pracht, P.; Bauer, C. A.; Grimme, S., Automated and efficient quantum chemical determination and energetic ranking of molecular protonation sites. *Journal of Computational Chemistry* **2017,** *38* (30), 2618-2631.

24. Case, D. A.; Cheatham Iii, T. E.; Darden, T.; Gohlke, H.; Luo, R.; Merz Jr, K. M.; Onufriev, A.; Simmerling, C.; Wang, B.; Woods, R. J., The Amber biomolecular simulation programs. *Journal of Computational Chemistry* **2005,** *26* (16), 1668-1688.

25. Aprà, E.; Bylaska, E.; de Jong, W.; Govind, N.; Kowalski, K.; Straatsma, T.; Valiev, M.; van Dam, H.; Alexeev, Y.; Anchell, J.; Anisimov, V.; Aquino, F.; Atta-Fynn, R.; Autschbach, J.; Bauman, N.; Becca, J.; Bernholdt, D.; Bhaskaran-Nair, K.; Bogatko, S., *NWChem: Past, Present, and Future*. 2020.

26. Pritchard, B. P.; Altarawy, D.; Didier, B.; Gibsom, T. D.; Windus, T. L., A New Basis Set Exchange: An Open, Up-to-date Resource for the Molecular Sciences Community. *J. Chem. Inf. Model.* **2019,** *59*, 4814-4820.

27. Wolken, J. K.; Tureček, F. e., Proton affinity of uracil. A computational study of protonation sites. *Journal of the American Society for Mass Spectrometry* **2000,** *11* (12), 1065-1071.

28. Nielson, F. F.; Colby, S. M.; Thomas, D. G.; Renslow, R. S.; Metz, T. O., Exploring the Impacts of Conformer Selection Methods on Ion Mobility Collision Cross Section Predictions. *Analytical Chemistry* **2021,** *93* (8), 3830-3838.

29. Reading, E.; Munoz-Muriedas, J.; Roberts, A. D.; Dear, G. J.; Robinson, C. V.; Beaumont, C., Elucidation of Drug Metabolite Structural Isomers Using Molecular Modeling Coupled with Ion Mobility Mass Spectrometry. *Analytical Chemistry* **2016,** *88* (4), 2273-2280.